\documentclass[aps,prb,12pt,preprint,amssymb,superscriptaddress,showpacs]{revtex4-1}

\usepackage{verbatim}
\usepackage[english]{babel}
\usepackage{amsmath,amsthm}
\usepackage{graphicx}
\usepackage{subfigure}
\usepackage{epsfig}

\begin{document}
\title{\Large{The role of oxygen vacancies on the structure and the density of states of iron doped zirconia.}}
\author{Davide Sangalli}
\affiliation{Laboratorio MDM - IMM - CNR via C. Olivetti, 2 I-20864 Agrate Brianza (MB) Italy}

\author{Alessio Lamperti}
\affiliation{Laboratorio MDM - IMM - CNR via C. Olivetti, 2 I-20864 Agrate Brianza (MB) Italy}

\author{Elena Cianci}
\affiliation{Laboratorio MDM - IMM - CNR via C. Olivetti, 2 I-20864 Agrate Brianza (MB) Italy}

\author{Roberta Ciprian}
\affiliation{Laboratorio MDM - IMM - CNR via C. Olivetti, 2 I-20864 Agrate Brianza (MB) Italy}

\author{Michele Perego}
\affiliation{Laboratorio MDM - IMM - CNR via C. Olivetti, 2 I-20864 Agrate Brianza (MB) Italy}

\author{Alberto Debernardi}
\affiliation{Laboratorio MDM - IMM - CNR via C. Olivetti, 2 I-20864 Agrate Brianza (MB) Italy}

\date{\today}
\begin{abstract}
In this paper we study, both with theoretical and experimental approach, the effect of iron
doping in zirconia. Combining density functional theory (DFT) simulations 
with the experimental characterization of thin films, we show that iron is in the
$Fe^{3+}$ oxidation state and accordingly that the films are rich in
oxygen vacancies ($V_O^{^{\bullet\bullet}}$).
$V_O^{^{\bullet\bullet}}$ favor the formation of the tetragonal phase in doped zirconia ($ZrO_2$:$Fe$)
and affect the density of state at the Fermi level as well as the local magnetization of $Fe$ atoms.
We also show that the $Fe(2p)$ and $Fe(3p)$ energy levels can be used as a marker for the
presence of vacancies in the doped system. In particular the computed position of the $Fe(3p)$ peak
is strongly sensitive to the $V_O^{^{\bullet\bullet}}$ to $Fe$ atoms ratio.
A comparison of the theoretical and experimental $Fe(3p)$ peak position suggests that in our films
this ratio is close to $0.5$.

Besides the interest in the material by itself, $ZrO_2$:$Fe$ constitutes a test case for the
application of DFT on transition metals embedded in oxides.
In $ZrO_2$:$Fe$ the inclusion of the Hubbard $U$ correction significantly changes the
electronic properties of the system. However the inclusion of this correction, at least for the
value $U=3.3\ eV$ chosen in the present work, worsen the agreement with the measured photo--emission
valence band spectra.
\end{abstract}
\maketitle

\section*{Introduction}
In dilute magnetic semiconductors (DMS) magnetic impurities, usually transition metals (TM),
are introduced to produce a magnetic ground state. These systems have been extensively investigated
since the discovery of carrier induced ferro--magnetism in $(In,Mn)As$~\cite{Ohno1992} and
$(Ga,Mn)As$~\cite{Ohno1996}, and are believed to be fundamental to fabricate
spin--based electronic devices.
The understanding of DMS physical properties constitutes a challenge for the theory as
the fundamental mechanism leading to ferromagnetic interaction can be hardly
explained~\cite{Coey2005_Nature}. Also experimentally the inclusion and the influence of TM
doping is not clearly understood.
Indeed, while several DMS were predicted to have a Curie temperature ($T_c$) above room temperature,
no experimental report of $T_c>300 K$ has been left unchallenged by other studies~\cite{Sato2010}.
Moreover some results suggest that magnetic impurities, at least at very low doping concentration, act
as paramagnetic centers~\cite{Gunnlaugsson2010}.
Recently a new class of DMS, based on oxides such as zirconia ($ZrO_2$) and hafnia ($HfO_2$),
has received great attention, after the experimental reports of room temperature
magnetism in $Fe$ doped $HfO_2$ and
$ZrO_2$~\cite{Hong2005,Hong2006,Hong2012,Coey2005,Kriventsova2006,Sahoo2009}
and the theoretical prediction of high $T_c$ in TM doped $ZrO_2$~\cite{Ostanin2007,Archer2007}.

For a better understanding of the magnetic properties of the system, a clear picture of its structural
and electronic properties is fundamental. As opposite to standard bulk materials,
where usually the most stable configuration can be unequivocally identified, in DMS the inclusion
of the dopant can induce stress, disorder and defects in the system with many possible
configurations close in energy.
From one side, theoretically, the modeling of the material, also at the first--principles levels,
requires some assumptions on the initial structure and on the position occupied by the dopant. From the
other side, experimentally, stress, disorder and defects make difficult to
provide a unique interpretation to the features observed.
Thus a combined approach is the best option.

Among the structural defects of dilute magnetic oxides (DMO), oxygen vacancies ($V_O^{^{\bullet\bullet}}$)
are believed to affect the magnetism~\cite{Hong2005,Coey2005,Hong2005_PRB}.
Indeed it has been suggested that delocalized electrons, associated with $V_O^{^{\bullet\bullet}}$,
can play a crucial role in the magnetization mechanisms of DMO~\cite{Coey2005_Nature}.
However, in this model, $V_O^{^{\bullet\bullet}}$ are assumed to always
induce delocalized states, which can mediate the magnetic interaction. This assumption
is true in the undoped oxide, while in presence of doping should be verified case by case.

In the present paper we describe the structural and electronic properties of iron
doped zirconia ($ZrO_2$:$Fe$) focusing our attention on the role of $V_O^{^{\bullet\bullet}}$ and
on their relation with the dopant.
The role of $V_O^{^{\bullet\bullet}}$ in $ZrO_2$:$Fe$,
and more in general of $ZrO_2$ doped with valence $+3$ elements ($X^{+3}$, with $X=Fe,Y,$ etc...)
has been, in part, explored in view of different applications, for
oxygen sensing~\cite{Cao2001,Cao2002,Stefanic2000}
and more recently for resistive switching memories~\cite{Zhang2010,Spiga2012}.
For $ZrO_2$:$Fe$ in particular only few experimental reports
exist. Also for TM doped oxides in general, no systematic theoretical
description
of the relation between $V_O^{^{\bullet\bullet}}$ and doping exist. For example the $V_O^{^{\bullet\bullet}}$
formation energy, in presence of doping, is usually considered~\cite{Zhang2010} only for the
$V_O^{^{\bullet\bullet}}$ to dopant atoms ratio, $y_{V_O^{^{\bullet\bullet}}/X}$, equal to 1 and 
again $V_O^{^{\bullet\bullet}}$ are assumed to induce delocalized states which
could mediate the electron conduction in case of resistive switching,
regardless of the value of $y_{V_O^{^{\bullet\bullet}}/X}$.

Instead, in case of $X^{+3}$ elements, like iron, the most stable configuration is expected to have
$y_{V_O^{^{\bullet\bullet}}/X}=0.5$ for charge compensation~\cite{Stapper1999}.
We thus focus our attention on this configuration describing how the properties of the system 
would change if $y_{V_O^{^{\bullet\bullet}}/X}$ deviates from the value $0.5$.

In sec.~\ref{sec:Framework} we describe both the theoretical and the
experimental approach to the description of
$ZrO_2$:$Fe$. The results from first--principles simulations are presented in sec.~\ref{sec:results_theory}.
The electronic and structural properties of the system are described in function of the doping
and oxygen vacancies concentration within
density functional theory (DFT) in the standard generalized gradient approximation (GGA).
For TM oxides the standard approximations to DFT are known to fail in the description of the
so called on--site correlation. Thus DFT can be corrected with a ``Hubbard'' term, DFT+$U$ scheme,
where $U$ is an external parameter, which improves the DOS of the valence electrons.
However little is known in the case of TM used as dopant in DMO. Thus we also
investigate how this term would influence the electronic properties of the system 
in $ZrO_2$:$Fe$.

The experimental results are then presented in sec.~\ref{sec:results_experiments}.
Here we show that, indeed, the measured properties best agrees with the
$y_{V_O^{^{\bullet\bullet}}/X}=0.5$ configuration.
Moreover a detailed comparison of the measured valence band (VB) and DFT
density of states (DOS) is done. This is a direct way to explore the value of
the on--site electronic correlation on this system, i.e. to adjust the value of the $U$
parameter to be used in the DFT+$U$ approch.

\section{Framework}\label{sec:Framework}
\subsection{Computational approach} \label{sec:Framework_TH}

We computed, from first--principles, the ground state of the two most common
phases of $ZrO_2$, i.e. the tetragonal and the monoclinic phases, at different doping concentrations.
We used the PWscf (4.3.2) package~\cite{QuantumEspresso}, considering a super--cell with 96 atoms
(few less when $V_O^{^{\bullet\bullet}}$ are considered) and in some cases also a smaller super--cell with
12 atoms for the description of the highest doping configuration. For all systems
the atomic positions are fully relaxed. The ground state was computed within 
the GGA~\cite{Perdew1996} to the DFT
scheme~\cite{Hohenberg1964,Kohn1965} with ultra--soft pseudo--potentials~\cite{Vanderbilt1990,Rappe1990}.
We used a 35 $Ry$ cut---off for the wave--functions, 400 $Ry$ cut---off for the augmentation density and
a Monkhorst--Pack grid 2x2x2 for the Brillouin zone to have the error on the energy differences 
between the monoclinic and the tetragonal phase lower than $1\ meV$ per formula unit ($f.u.$);
this was the most stringent condition for our simulations.
We estimated the error on the total energy to be lower than $0.1\ eV/f.u.$.
Convergence paramenters are $10^{-8}\ Ry$ on the total energy for the scf cycles
and both $10^{-4}\ Ry$ on the total energy and $10^{-3}\ Ry/Bohr$ on the forces for the atomic relaxation.
The pseudo--potential of $Zr$ includes semi--core electrons.
$Fe$ atoms were placed at the substitutional $Zr$ sites and kept as far as possible from each other
to mimic uniform doping. For $V_O^{^{\bullet\bullet}}$ instead we considered many different configurations
(see discussion in sec.~\ref{sec:results_theory}), specifically
we considered $ZrO_2$:$Fe$ at the atomic doping concentration $x_{Fe}=6.25\%,\ 12.5\%,\ 18.75\%,\ 25\%$
with, $y_{V_O^{^{\bullet\bullet}}/Fe}=0.5$, and without, $y_{V_O^{^{\bullet\bullet}}/Fe}=0$, oxygen vacancies.
We also considered $y_{V_O^{^{\bullet\bullet}}/Fe}=1.0$ for $x_{Fe}=6.25\%,\ 25\%$.
In total we studied about $50$ different systems of
$Zr_{1-x}Fe_xO_{2-z}V_{Oz}^{^{\bullet\bullet}}$ changing $x_{Fe}$ and
$z_{V_O^{^{\bullet\bullet}}}=x_{Fe}\times y_{V_O^{^{\bullet\bullet}}/Fe}$ for either the monoclinic
or the tetragonal structure.
For few selected configurations, i.e. at the lowest and the highest considered doping concentrations
$x_{Fe}=6.25,\ 25\%$, we also performed calculations within the
simplified GGA+$U$ approach~\cite{Cococcioni2005} implemented in the PWscf package, again 
considering $y_{V_O^{^{\bullet\bullet}}/Fe}=0,\ 0.5,\ 1$, in order to explore the effect of the 
Hubbard correction on the electronic structure of the system. The results are
presented mainly for the high--doping situation which we have also experimentally.
The configurations at $y_{V_O^{^{\bullet\bullet}}/Fe}=0$ and $1$ resulted to be metallic and in these
case the convergence of the physical quantities against the sampling of the k--points grid
was verified.

The cell parameters for both the tetragonal and the monoclinic phase of
pure $ZrO_2$ are the same used in Ref.~\onlinecite{Sangalli2011}. Specifically for the monoclinic
phase $a=5.18$ \AA, $b/a=1.011$, $c/a=1.037$ and $\beta=99^\circ 10'$; while for the tetragonal phase
$a=5.18$ \AA and $c/a=1.0305$. The same parameters were used
for $ZrO_2$:$Fe$ as well. However we even performed a full relaxation of
our 96 atoms super--cell for few selected configurations and we found out that this
have a negligible impact on the properties of the system here considered.

In sec.~\ref{sec:results_theory} we systematically compare the results of the present simulations
with the $ZrO_2$:$Y$ ($Y$ doped $ZrO_2$) system. Yttrium is one of the most studied and
used dopant of $ZrO_2$ and shares with iron the same valence. All the data reported for $ZrO_2$:$Y$
are from ref.~\onlinecite{Sangalli2011}.

In order to describe the semi--core levels of iron and compare the results with XPS
measurements, we run calculations with a norm--conserving fully--relativistic approach.
To this end, we used Hartwigsen, Goedecker, and Hutte (HGH) pseudo--potentials~\cite{Hartwigsen1998}
which contain semi--core electrons in valence and are constructed with a fully relativistic calculation.
The latter are not available within the PWscf~\cite{QuantumEspresso}
code and so we used the abinit (6.8) code~\cite{Abinit}.
We studied the semi--core levels only for the $x_{Fe}=25\%\ at.$ case again considering
$y_{V_O^{^{\bullet\bullet}}/Fe}=0,\ 0.5,\ 1$. 
We used smaller super--cells, 12 atoms ($y_{V_O^{^{\bullet\bullet}}/Fe}=0$ and $1$) and a 24 atoms supercell
($y_{V_O^{^{\bullet\bullet}}/Fe}=0.5$), with cut--off of 170 $Ry$ and a
Monkhorst--Pack grid 3x3x3 and 3x3x2 respectively for the Brillouin zone to have the error on the
energy levels position lower then $0.1\ eV$. The very high energy cut--off was needed, as 
the norm conserving HGH pseudo--potentials are harder than the ultra--soft ones used with PWscf
and also because the semi--core levels
are much more localized than valence electrons.
The value $x_{Fe}=25\%\ at.$ was chosen to have smaller super-cells but also because this is
quite close to the experimentally measured doping concentration in our films.
The atomic positions instead were obtained relaxing the same structures with the PWscf code and
then we checked that the residual forces on the atoms computed with Abinit were negligible.

Finally for a quantitative comparison of the measured photo--emission and
the computed valence band we have performed calculations within GGA+$U$ at $U=1.0,\ 2.0,\ 3.3\ eV$
at $x_{Fe}=18.75\%\ at.$ and $y_{V_O^{^{\bullet\bullet}}/Fe}=0.5$.
A theoretical smearing of $0.02\ Ry$ was used to generate the DOS used in
Figs.~\ref{Fig:DFT_DOS}-\ref{Fig:DFT+U_DOS} while a higher smearing
of $0.06\ Ry$ was used for the DOS in Fig.~\ref{Fig:XPS_and_DFT} to mimic the experimental
peak width.

\subsection{Experimental setup} \label{sec:Framework_Exp}

Experimentally
$ZrO_2$ and $ZrO_2$:$Fe$ thin films were grown on $Si/SiO_2$ substrates in a flow--type hot wall
atomic layer deposition reactor (ASM F120) starting from $\beta$--diketonates metalorganic
precursors, namely $Zr(C_{11}H_{19}O_2)_4$ for $Zr$
and $Fe(C_{11}H_{19}O_2)_3$ for $Fe$. To grant a stable reactivity, $Zr$ precursor was
kept at $170^\circ$C, while $Fe$ precursor was maintained at $115^\circ$C.
Ozone was used as oxidizing gas in the reaction process
The film growth was achieved by alternately introducing the reactants separated by $N_2$ inert
gas purging pulses. The $Fe$ concentration in $ZrO_2$:$Fe$ films was tuned tailoring the $Zr/Fe$
precursors pulsing ratio
and the growth temperature was maintained at $350^\circ$C (details in Ref.~\onlinecite{Lamperti2012}).
After the deposition the films were annealed at $600^\circ$C in $N_2$ flux for 60s.
The growth parameters were tuned in order to fix the thickness, $d=19\pm1\ nm$, and
the doping concentration $x_{Fe}=20\%\pm3\%$ for the $ZrO_2$:$Fe$ films.
$x_{Fe}$ was chosen in order to stabilize the tetragonal phase according to our
theoretical results.

Film crystallinity was checked by X--ray diffraction (XRD) at fixed grazing incidence angle $\omega=1^\circ$
and using $Cu$ $K_\alpha$ ($\lambda = 0.154\ nm$) monochromated and collimated X--ray beam
(details in Ref.~\onlinecite{Lamperti2011}).
Film uniform doping along its thickness was checked by Time of Flight Secondary Ion Mass Spectrometry (ToF--SIMS)
depth profiling using an ION--TOF IV instrument, with $500\ eV$ $Cs^+$ ions for sputtering and $25\ keV$ $Ga^+$ ions
for analysis. Secondary ions were collected in negative polarity and interlaced mode. Recorded intensities
were normalized to $^{30}Si$ intensity in bulk silicon. The instrument depth resolution
is below 1 nanometer.

To elucidate $Fe$ chemical state and concentration in $ZrO_2$:$Fe$
films, X--ray photo--emission (XPS) measurements were performed on a PHI 5600 instrument equipped
with a monochromatic Al $K_a$ x-ray source
($E = 1486.6$ eV) and a concentric hemispherical analyzer. The spectra were collected at a
take--off angle of $45^\circ$ and band--pass energy $11.50\ eV$. The instrument resolution is $0.5\ eV$.

\section{First principles predictions}\label{sec:results_theory}

\subsection{$V_O^{^{\bullet\bullet}}$ and structural properties}

\begin{figure}[t]
\includegraphics[width=0.9\textwidth]{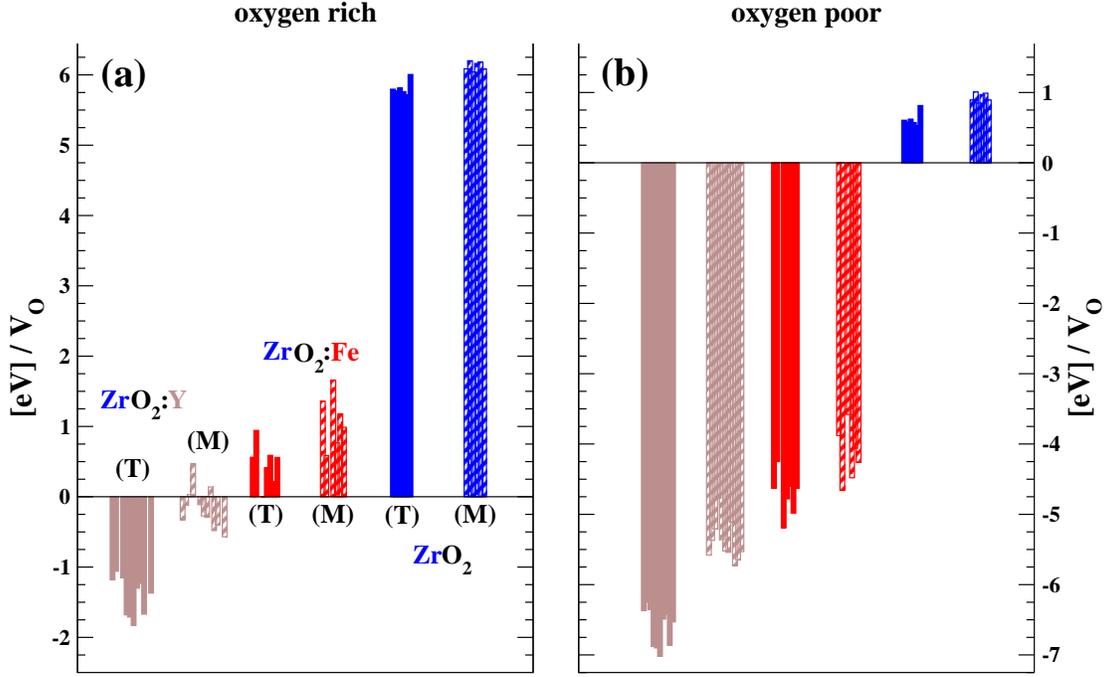}
\caption{(color online)
DFT (GGA) formation energy of oxygen vacancies (see Eq.~\ref{eq:Vo_Formation_E})
in $ZrO_2$:$Y$, $ZrO_2$:$Fe$ and
$ZrO_2$ in the two extrema case of (a) oxygen rich conditions and (b) oxygen poor conditions
in both the tetragonal and the monoclinic structure. The doped systems are considered in the
charge compensated configuration (i.e. $y_{V_O^{^{\bullet\bullet}}/X}=0.5$ for $X=Fe,Y$).
The values are computed for different oxygen vacancies concentrations and also varying, for some
concentrations, the position of the oxygen vacancies. In panels $(a)$ and $(b)$ histograms are presented
in the same order (and colors).
}
\label{Fig:DFT_Vo}
\end{figure}

In the literature $ZrO_2$:$Fe$ has been studied as a candidate material for oxygen sensing applications
because $Fe^{+3}$ atoms, replacing  $Zr^{+4}$ atoms, are expected
to induce oxygen vacancies for charge compensation~\cite{Cao2001,Cao2002}.
Thus, as a first step, we consider the 
$V_O^{^{\bullet\bullet}}$ formation energy:
\begin{equation} \label{eq:Vo_Formation_E}
\Delta E_1(x_{Fe},z)=\big(\ E[Zr_{1-x}Fe_xO_2]
-(\ E[Zr_{1-x}Fe_xO_{2-z}]+(z/2)\mu[O_2]\ ) \ \big)/z
\end{equation}
at fixed $y_{V_O^{^{\bullet\bullet}}/Fe}=z/x=0.5$, i.e. for a charge compensated system.
Here we considered both the oxygen rich condition (Fig.~\ref{Fig:DFT_Vo}.(a),
$\mu[O_2]=E[O_2]$ with $E[O_2]$ the total energy of an isolated
oxygen molecule in its ground state)
and the oxygen poor condition (Fig.~\ref{Fig:DFT_Vo}.(b),
$\mu[O_2]=E[ZrO_2]-E[Zr]$).
The formation energy for $ZrO_2$:$Fe$ is compared with the case of pure 
$ZrO_2$, $\Delta E_1(0,z)$
and $ZrO_2$:$Y$, $\Delta E_1(x_Y,z)$.
To this end we considered different
$V_O^{^{\bullet\bullet}}$ concentrations and, for each,
different $V_O^{^{\bullet\bullet}}$ configurations.
However we found that $\Delta E(x,z)$ is mainly determined by the kind of dopant,
while the influence of the other parameters is lower. In Fig.~\ref{Fig:DFT_Vo} the changes due of
these parameters results in different values for each system.
While the $V_O$ formation energy is negative in $ZrO_2$:$Y$ already in the oxygen rich case,
in $ZrO_2$:$Fe$ films it is slightly positive, i.e. $\Delta E_1^{tetra}\approx 0.5\ eV$,
but ten times lower than in pure $ZrO_2$. Varying the chemical potential from the oxygen rich to the 
oxygen poor configuration $\Delta E_1^{tetra}$ becomes negative, thus $Fe$ favors the formation
of $V_O^{^{\bullet\bullet}}$.

\begin{figure}[t]
\includegraphics[width=0.8\textwidth]{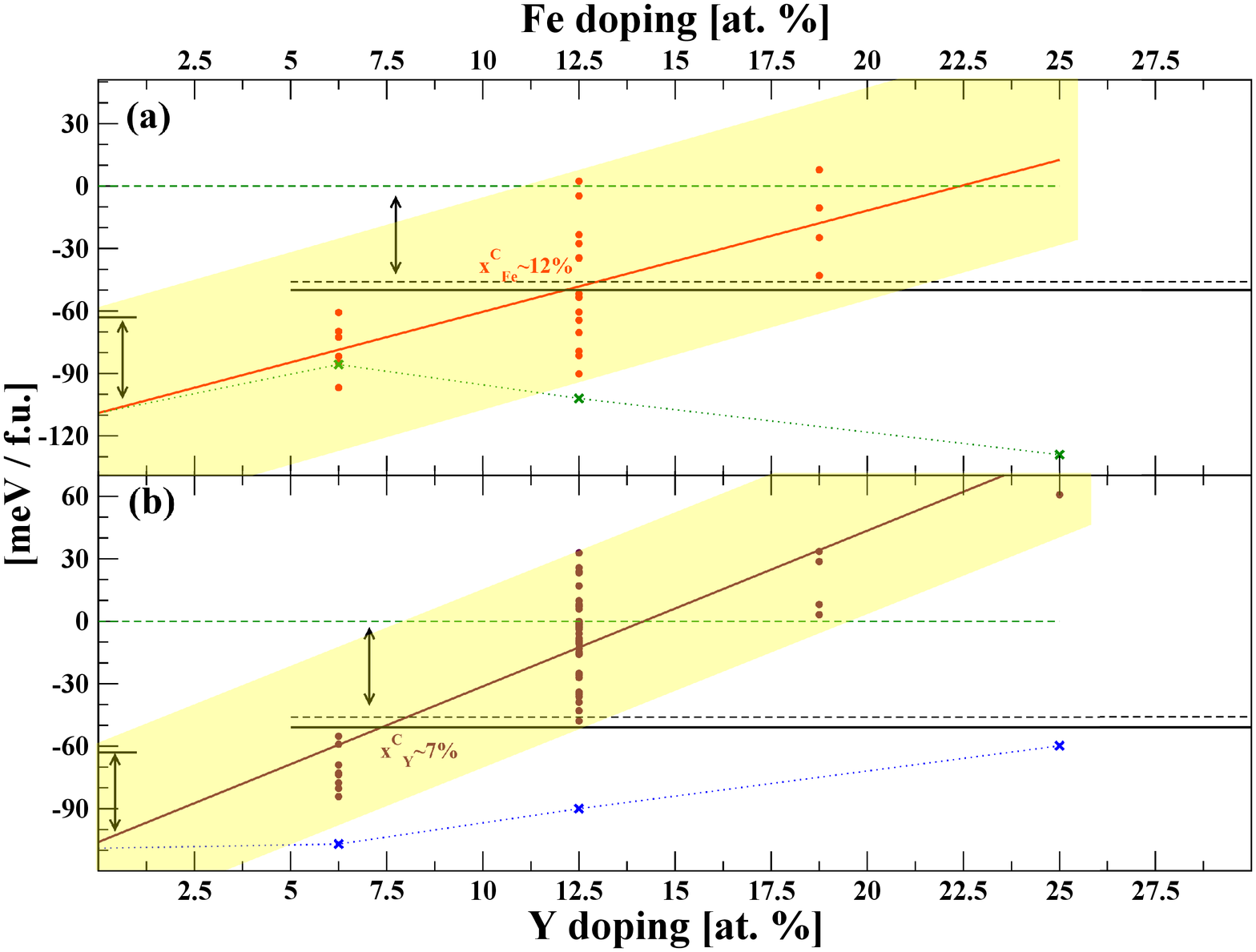}
\caption{(color online)
DFT (GGA) total energy difference per formula unit
between the tetragonal against the monoclinic phase for $ZrO_2$:$Fe$,
panel (a), and $ZrO_2$:$Y$, panel (b).
Total energies are computed for the charge compensated system (dots,
$y_{V_O^{^{\bullet\bullet}}/X}=0.5$ for $X=Fe,Y$) changing the atomic configurations
for each given concentrations. The shadowed areas are guides for the eyes while the continuous
lines are a linear fit of the data. Also the results for the systems without oxygen vacancies
(crosses, $y_{V_O^{^{\bullet\bullet}}/X}=0$) are shown for comparison.
The zero level is shifted of (i) $-46\ meV/f.u.$ to align the energy difference
at zero doping with the experimental value,
(ii) $-5\ meV/f.u.$ to include the computed zero--point--energy difference of
the two lattices.}
\label{Fig:DFT_TM}
\end{figure}

The creation of oxygen vacancies induces disorder in the system (see also the inset in Fig.~\ref{Fig:XRD})
thus the most symmetric phases are expected to be favored against the monoclinic phase.
To evaluate this effect in 
Fig.~\ref{Fig:DFT_TM} we consider the energy difference $\Delta E_2(x)$
between the tetragonal and the monoclinic phase as a function of the doping concentration,
at fixed $y_{V_O^{^{\bullet\bullet}}/X}=0.5$.
We look for the iron atomic percent, $x^C_{Fe}$, at which the tetragonal phase becomes favored.

The value of $\Delta E_2$ is very small and thus at the limit of the DFT--GGA resolution.
The computed energy difference between the two phases at zero doping is
${\Delta E_2(0)=109\ meV/f.u.}$, in agreement with previous works,
($63\ meV/f.u.$\cite{Stapper1999}, $144\ meV/f.u.$\cite{Dewhurst1998}); the experimental estimation
is $63\ meV/f.u.$\cite{Achermann1975}.
It is reasonable to assume that the trend of the energy difference is better computed than its
absolute value and accordingly, assuming a constant ``zero--doping error'' of $\approx46\ meV/f.u.$ for every $Fe$
concentration, we can subtract it.
Being ${\Delta E_2}$ of the order of few $meV/f.u.$ also the phonon energy of the two lattice
could play a role. Indeed the monoclinic to tetragonal phase transition at $\approx 1440\ K$ can be 
explained in this terms~\cite{Lou2009,Debernardi}. Thus we considered the energy difference
of the lattice between the two structures for the undoped system.
At room temperature however 
we found this contribution to be almost negligible, $\approx 5\ meV/f.u.$.

$\Delta E_2$ come out, instead, to be particularly sensitive to the chosen atomic configuration.
Accordingly the data in Fig.~\ref{Fig:DFT_TM} are scattered,
with $\Delta E_2$ changing of few $meV/f.u.$ at given $x_{Fe}$.
To extract the exact $x^C_{Fe}$ a statistical
occupation of the different configurations should be considered. However, to this end,
one should sample a huge number of configurations, which is not feasible within DFT.
In the present paper we assumed that, fixed $x_{Fe}$ and $z_{V_O^{^{\bullet\bullet}}}$,
changing the configurations for the $V_O^{^{\bullet\bullet}}$,
$\Delta E_2$ spans uniformly a given
energy range ($E_R$) which can be extrapolated considering a limited number of configurations.
$E_R$ is expected to increase, increasing the doping concentration,
as an increasing number of configurations becomes available.
With this assumptions $x^C_{Fe}$ was extracted considering the central value of the $E_R$.

In practice this was done with a linear fit of the data.
In Fig.~\ref{Fig:DFT_TM}, to obtain the critical doping concentration,
$\Delta E_2$ at zero doping is matched at the experimental value $63\ meV$,
while the theoretical results would be $109\ meV$.
The result, $x^C_{Fe}\approx12\%\ at.$,
can be compared with the case of $ZrO_2$:$Y$, where the same approach gives $x^C_{Y}\approx7\%\ at.$
which exactly matches the experimental value~\cite{Sangalli2011}.
We stress that with this approach the exact doping concentration can be affected by an error which
can be as large as few atomic percent. What is significant here is the comparison of the
two systems, i.e. $ZrO_2$:$Y$ and $ZrO_2$:$Fe$.
Indeed both dopants, inducing oxygen vacancies favor the tetragonal against the monoclinic structure.
However the two linear fits posses different slopes and we can conclude that iron is less efficient
than yttrium in inducing a monoclinic to tetragonal phase transition.
In Fig.~\ref{Fig:DFT_TM} we also report the energy difference between the monoclinic and the tetragonal
phase for the case without oxygen vacancies, i.e. $y_{V_O^{^{\bullet\bullet}}/X}=0$. In this configuration
we found that the local structure of the crystal is much less distorted by doping
and accordingly the variation of the energy difference between the two phases is small.
This confirms that a key role in the monoclinic to tetragonal phase transition is played
by oxygen vacancies~\cite{Sangalli2011} and not by the dopant itself.

\subsection{Electronic properties}

Given the results of the previous section and the fact that experimentally we describe a system at high 
doping concentration, which we found to be in the tetragonal phase, in the description of the electonic
properties of the system we focus our attention on the tetragonal structure of $ZrO_2$:$Fe$.

\begin{figure}[t]
\includegraphics[height=0.6\textheight]{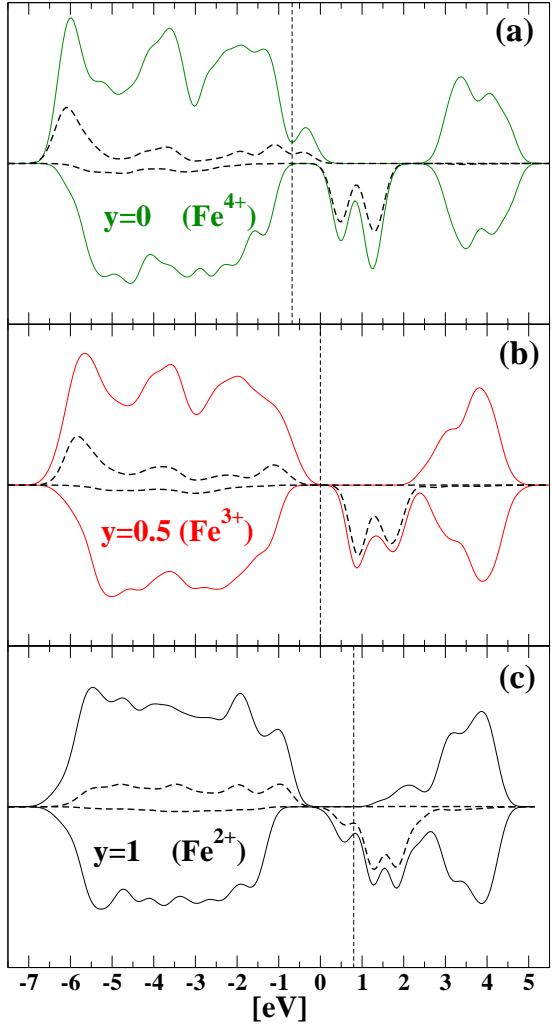}
\caption{(color online)
Total (full line) and $d$--orbital projected (dashed line) density of states (DOS) at
the GGA level of $ZrO_2$:$Fe$ at $x_{Fe}=25\%$ with
$y_{V_O^{^{\bullet\bullet}}/Fe}$ equal to respectively $0$ (panel a), $0.5$ (panel b), $1$ (panel c).
The vertical dashed line marks the Fermi level. The Fermi level of panel (b) is the zero of the energy
axis, while in panels (a) and (c) the zero is obtained aligning the bottom of the valence band at $\approx-6.5\ eV$
as in panel (b). }
\label{Fig:DFT_DOS}
\end{figure}

The main difference between $Y$ and $Fe$ is
the presence of the unfilled $Fe(d)$ orbitals which, falling
inside the energy gap of zirconia, determine the electronic properties of the doped
system. The $d$--orbitals occupation is also strongly affected by $V_O^{^{\bullet\bullet}}$
and is used here, together with the computed magnetic moment, to infer the $Fe$ oxidation state.
At $y_{V_O^{^{\bullet\bullet}}/Fe}=0$, $Fe$ acts as an acceptor (Fig.~\ref{Fig:DFT_DOS}.(a))
with the creation of holes in the majority spin VB. These are
preferentially located on the $Fe(d)$ orbitals as shown by the projected--DOS
with the projection of the hole states on the $d$--orbitals close to 0.5.
Thus $Fe$ is forced in the $Fe^{+4}$ oxidation state with a magnetic moment
per atom equal to 4 Bohr magnetons ($\mu_B$).
The creation of $V_O^{^{\bullet\bullet}}$ release
the electrons captured by the $O$ anions.
At $y_{V_O^{^{\bullet\bullet}}/Fe}=0.5$ the
system turns into a charge--transfer semi--conductor (see Fig.~\ref{Fig:DFT_DOS}.(b)),
i.e. the $V_O^{^{\bullet\bullet}}$ do not create an impurity bands, as it would happen in $ZrO_2$,
but compensate the holes in the $Fe(d)$ orbitals.
In this configuration $Fe$ atoms are in the $+3$ oxidation state and 
the magnetic moment per iron atom is maximized, 5 $\mu_B$.
If $y_{V_O^{^{\bullet\bullet}}/Fe}$ exceeds 0.5, electrons
start to fill the minority $Fe(d)$ levels. This decreases the average
magnetic moment, while the system reverts to an half--metal.
At $y_{V_O^{^{\bullet\bullet}}/Fe}=1$ (Fig.~\ref{Fig:DFT_DOS}.(c)) all iron atoms are in a
$+2$ oxidation state with the per atom magnetic moment equal to 4 $\mu_B$.
In Fig.~\ref{Fig:DFT_DOS} we also notice that at
$y_{V_O^{^{\bullet\bullet}}/Fe}\le 0.5$ no extra state, other than the $Fe(d)$ orbitals, appears
between the valence and the conduction band of $ZrO_2$. Only when $y_{V_O^{^{\bullet\bullet}}/Fe}> 0.5$
(Fig.~\ref{Fig:DFT_DOS}.(c)) such a state exists. The latter can be associated
to an impurity band which
has been suggested to create bound magnetic polarons in case of magnetic doping~\cite{Coey2005_Nature}.
However the configuration $y_{V_O^{^{\bullet\bullet}}/Fe}=1$ is not favored.
Indeed the energy cost, for each extra $V_O^{^{\bullet\bullet}}$ created in the system, of the reaction
\begin{equation}
Zr_{1-x}O_{2-x/2}Fe_x\ \rightarrow\ Zr_{1-x}O_{2-x}Fe_x +\frac{x}{4}\mu[O_2]
\end{equation}
changes from $\approx 2.5\ eV$, oxygen rich conditions, to $\approx 0.\ eV$, oxygen poor conditions, thus
remaining positive for any value of the oxygen chemical potential.
As for the $V_O^{^{\bullet\bullet}}$ formation energy $\Delta E_1$, this value is weakly dependent on the
atomic doping $x_{Fe}$.
Last but not least, even if in this case the impurity band exist, it is empty.
Thus the possible existence of bound magnetic polaron in $ZrO_2$:$Fe$ is unlike.
We will also show in the next section that, experimentally, iron in $ZrO_2$:$Fe$
is in the $Fe^{+3}$ and not in the $Fe^{+2}$ oxidation state.

We remark that, even if at $y_{V_O^{^{\bullet\bullet}}/Fe}=0$ and $y_{V_O^{^{\bullet\bullet}}/Fe}=1$
the system is metallic, the per atom magnetic moment is integer. The reason is that in both cases 
$ZrO_2$:$Fe$ is indeed an half--metal and thus electrons can move across the Fermi level only
in one spin channel.
We have verified this result increasing the sampling of the k--points grid from $2x2x2$ to $3x3x3$ in
the 96 super--cell and from $4x4x4$ to $8x8x8$ in the 12 atoms super--cell. In both cases the system
remains metallic, with fractional occupation in the majority ($y_{V_O^{^{\bullet\bullet}}/Fe}=0$)
or minority ($y_{V_O^{^{\bullet\bullet}}/Fe}=1$) spin channel (a smearing of $0.002\ Ry$ was used
in the self--consisten cycle in this case), but with constant per atom magnetic moment $m_z= 4\ \mu_B$.
In principle the 96 atoms super--cell with a sampling $3x3x3$ is equivalent to the 12 atoms supercell
with sampling $6x6x6$. However the two could differ because in the 96 atoms super--cell,
removing symmetries, disorder is taken into account. This could for example induce a localization of 
holes on the $Fe$ atoms. Thus the convergence check were also a rough way to explore
possible Anderson--like localization mechanisms. However we did not observe such phenomena.

\begin{figure}[t]
\includegraphics[height=0.6\textheight]{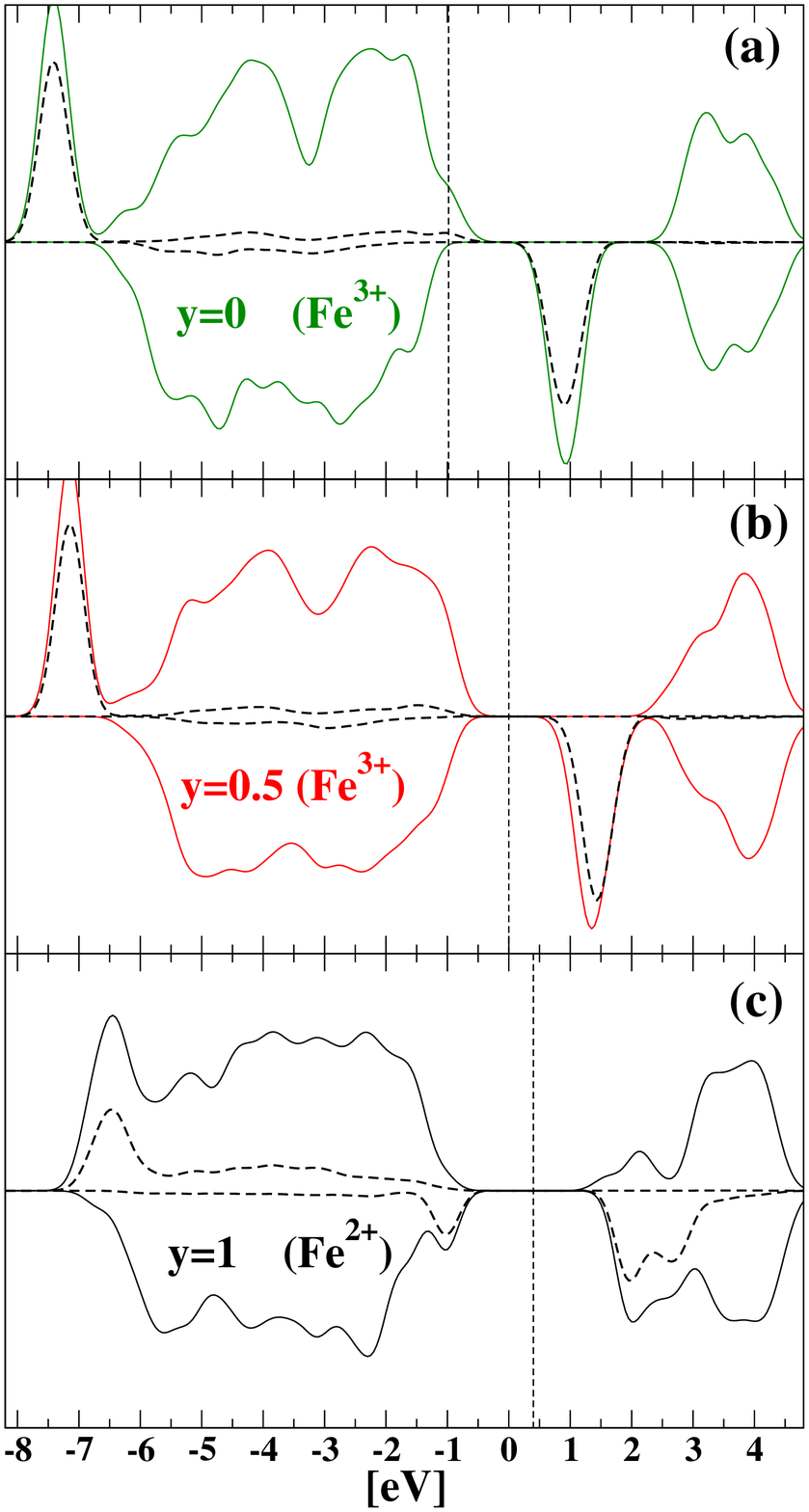}
\caption{(color online)
Total (full line) and $d$--orbital projected (dashed line) density of states (DOS) at
the GGA+$U$ level of $ZrO_2$:$Fe$ at $x_{Fe}=25\%$ with
$y_{V_O^{^{\bullet\bullet}}/Fe}$ equal to respectively $0$ (panel a), $0.5$ (panel b), $1$ (panel c).
The vertical dashed line marks the Fermi level. The Fermi level of panel (b) is the zero of the energy
axis, while in panels (a) and (c) the zero is obtained aligning the top of the conduction band at $\approx5\ eV$
as in panel (b). }
\label{Fig:DFT+U_DOS}
\end{figure}

These are the prediction of the GGA. Howerver
for TM oxides this approximation is known to suffer of some deficiencies. In particular it suffers
of the well known problem of self--interaction, which tends to delocalize too much the $d$ orbitals.
A common way to avoid this problem is to correct the DFT scheme with a Hubbard like term $U$ which
enters as an external parameter.
The value of $U$ is system dependent and should be optimized either with a direct comparison with
experimental data or with a self--consistent approach.
In the literature usually $U=1-3\ eV$ for elemental iron and $U=2-6\ eV$ in iron oxides.
For example Cococcini et al.~\cite{Cococcioni2005} report, after a self--consistent calculation, $U\approx 2.2\ eV$
for metal iron and $U=4.3\ eV$ for $FeO$. Here we begin choosing an intermediate value,
$U=3.3\ eV$, in order to evaluate the physical effects introduced by this
correction. 

In Fig.~\ref{Fig:DFT+U_DOS} we plot the DOS for the GGA+$U$ approach at $y_{V_O^{^{\bullet\bullet}}/Fe}=0,\ 0.5,\ 1$.
We can directly compare the results with the GGA DOS plots in Fig.~\ref{Fig:DFT_DOS}. As expected the $U$ correction
pushed down the occupied $d$ level and a sharp structure appeared in the DOS just below the VB of $ZrO_2$:$Fe$.
Also the crystal field splitting of the spin minority $d$ orbitals, between the $e_g$ and the $t_g$ states,
was reduced, and is not distinguishable anymore with the smearing parameter used in the plot;
with the exception of the case $y_{V_O^{^{\bullet\bullet}}/Fe}=1$.
However in the charge--compensated situation, $y_{V_O^{^{\bullet\bullet}}/Fe}=0.5$, these corrections
do not alter the qualitative description of the 
system, which remains a magnetic semi--conductor with the magnetic moment per atom maximized.
Instead, when we deviate from this configuration, we notice two important differences.
For $y_{V_O^{^{\bullet\bullet}}/Fe}<0.5$ the holes created in the VB are less localized
on the $Fe$ atoms. Indeed the projection of the hole states on the $d$--levels drops from $\approx 0.5$ (GGA)
to less than $0.1$ (GGA+$U$). Thus iron is in the $Fe^{3+}$ configuration, while
the holes are in the $ZrO_2$ VB, i.e. on the oxygen atoms. Accordingly the $V_O^{^{\bullet\bullet}}$
formation energy drops from $~0.5\ eV$ (GGA) to $~0.0\ eV$ because oxygen atoms are more weakly
bound to the system. 
For $y_{V_O^{^{\bullet\bullet}}/Fe}>0.5$ the extra electrons start to fill
the minority $d$-levels, as in the GGA case. However the newly occupied levels are pushed down in energy and thus the system
is not metallic but it displays an energy gap, i.e. GGA+$U$ predicts a a Mott insulator in this case.
Also for the $GGA+U$ case we verified that in the metallic case (i.e. at $y_{V_O^{^{\bullet\bullet}}/Fe}=0$)
the value of the magnetic moment remains constant improving the sampling of the Brillouin zone.

The electronic properties in the present section were reported for $x_{Fe}=25\%$. We did not
find significant changes for the other doping concentrations, at least for $y_{V_O^{^{\bullet\bullet}}/Fe}=0.5$.
At the lowest computed doping concentration however, $x_{Fe}=6.25\%$, the $Fe$ atoms are
too far apart and the localized $d$--levels do not create a band. Thus the metallic phases predicted 
within GGA ($y_{V_O^{^{\bullet\bullet}}/Fe}=0,\ 1$) become semi--conducting phases with defects states
localized close to the Fermi level.

\section{Experimental results}\label{sec:results_experiments}

\subsection{Structural characterization}

\begin{figure}[t]
\includegraphics[width=0.6\textwidth]{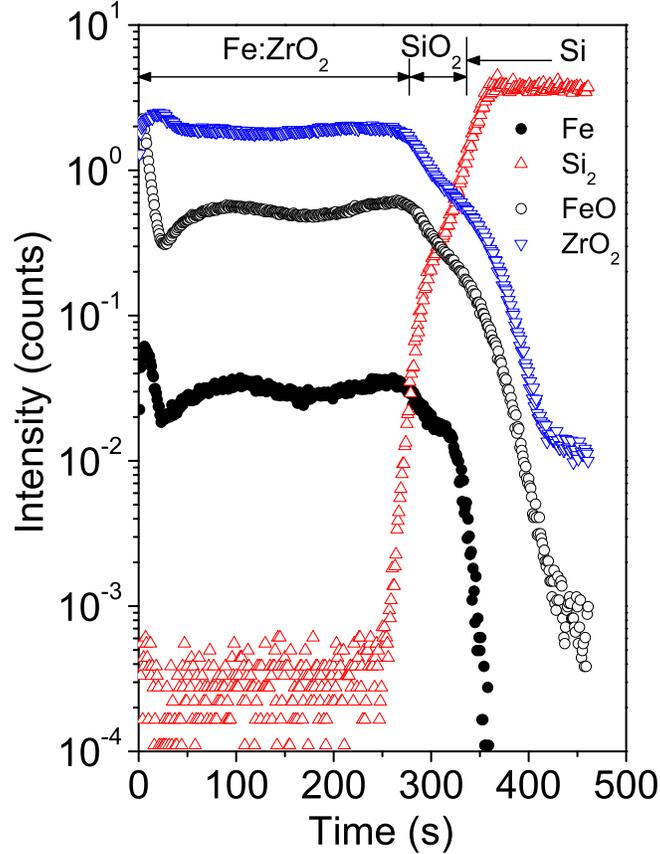}
\caption{(color online) Tof--SIMS depth profile of
$ZrO_2$:$Fe$ at $x_{Fe}\approx20\%\ at.$ .}
\label{Fig:TOF}
\end{figure}

Experimentally, as a first step, we first studied the structural properties of the films growth
by atomic layer deposition.
In Fig.~\ref{Fig:TOF} the ToF--SIMS
depth profile of a representative film (namely, $ZrO_2$:$Fe$ at $x_{Fe} = 20\%\ at.$),
including $Fe$, $FeO$, $ZrO$ and $Si$ negative secondary ion intensity profile is graphed.
$Fe$ and $FeO$ are both used as representative of $Fe$ distribution along the film thickness; in particular
$FeO$ ion fragment has not to be considered as a mark of $FeO$ chemical compound in the film, but as a fingerprint
of $Fe$ embedded in the $ZrO_2$ host matrix. The flatness of $ZrO$ and $Fe$ related profiles
indicates that the film grows uniformly during the ALD process, without changes in the distribution of the
chemical species, evidencing that the growth process is well controlled. Further, Si diffusion in $ZrO_2$
is excluded with a well distinct film/substrate interface, an indication that the substrate does not affect $ZrO_2$:$Fe$
properties both during the film growth and the thermal treatment. Furthermore,
the $Fe$ profile is almost constant, thus it is the doping in the film, and the absence of large fluctuations
such as peaked maxima, can exclude $Fe$ clustering.
Indeed the latter would have been observed as a sudden increase of
$Fe$ intensity with a concomitant abrupt decrease of $FeO$ intensity, indicating that an $Fe$ rich /
$O$ poor environment is detected. Instead
both $Fe$ and $FeO$ signals mimic the same profile shape, confirming that $Fe$ is uniformly diluted
within the $ZrO_2$ matrix.

\begin{figure}[t]
\includegraphics[width=0.85\textwidth]{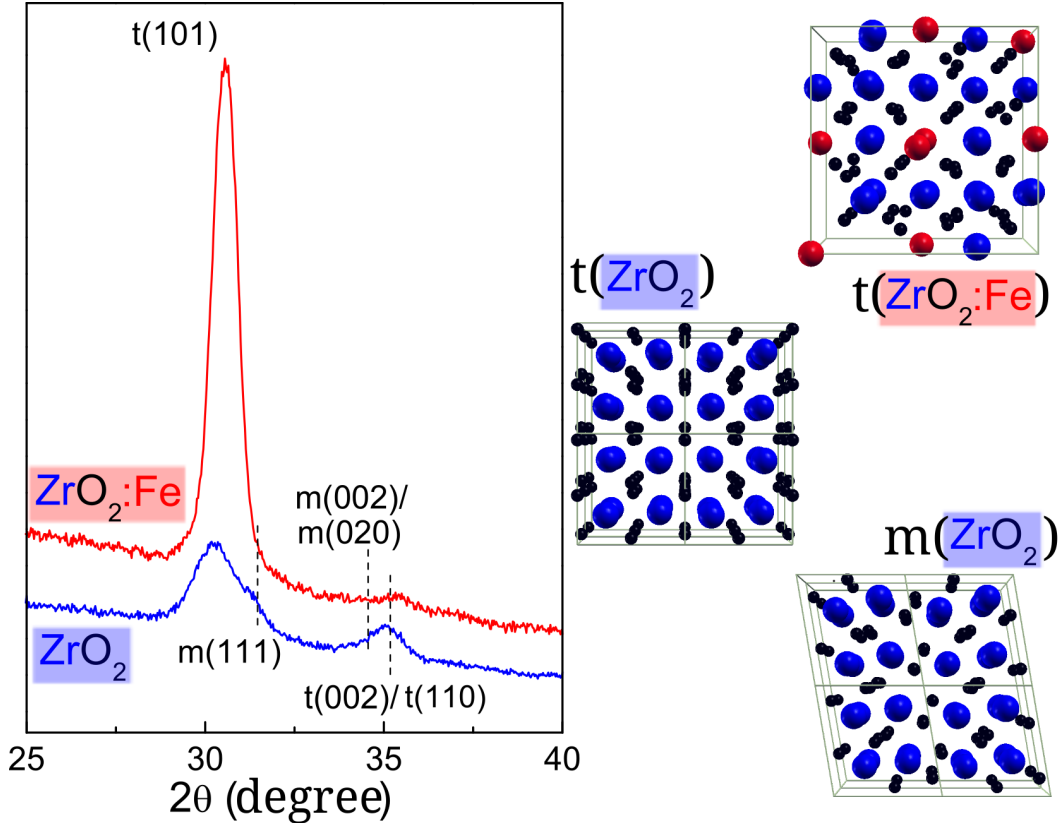}
\caption{(color online) - XRD patterns of $ZrO_2$ (blue) and
$ZrO_2$:$Fe$ (red) ($Fe$ doping ${\approx20\%\ at.}$)
films evidencing Fe doping is effective in suppressing the monoclinic phase.
t($ZrO_2$) and m($ZrO_2$) indicates the reflections from reference tetragonal and monoclinic
$ZrO_2$, respectively~\cite{XRD_structure}. On the right the relaxed DFT structure for
m($ZrO_2$), t($ZrO_2$) and t($ZrO_2$:$Fe$) at $x_{Fe}=25\%\ at.$ represented with the
xcrysden package (see Ref.~\onlinecite{xcrysden});
$Zr$ atoms in blue, $Fe$ atoms in red and the smaller $O$ atoms in black.}
\label{Fig:XRD}
\end{figure}

To get details on the film crystalline structure,
in Fig.~\ref{Fig:XRD} we compare the XRD patterns of $ZrO_2$ and $ZrO_2$:$Fe$. Both films mainly
present the cubic/tetragonal phase. Indeed in these films there is a balance between the
bulk energy, where the monoclinic phase is favored, and the surface energy, where the tetragonal phase
is favored.
The critical grain size~\cite{Christensen1998,Cerrato1997} below which the tetragonal phase
become the most favored is $\approx 15\ nm$.
In our films, being the grain size close to the film thickness (from XRD data),
we are close to this critical value. This can be evinced from the XRD patterns of pure $ZrO_2$ where
the peaks of the monoclinic phase are also evident. However in the $ZrO_2$:$Fe$ films the monoclinic
phase is completely suppressed, confirming our theoretical findings.
Even from these measures there is no indication
of segregated iron phase or iron oxide clusters.

\subsection{Electronic properties}

\begin{figure}[t]
\includegraphics[width=0.9\textwidth]{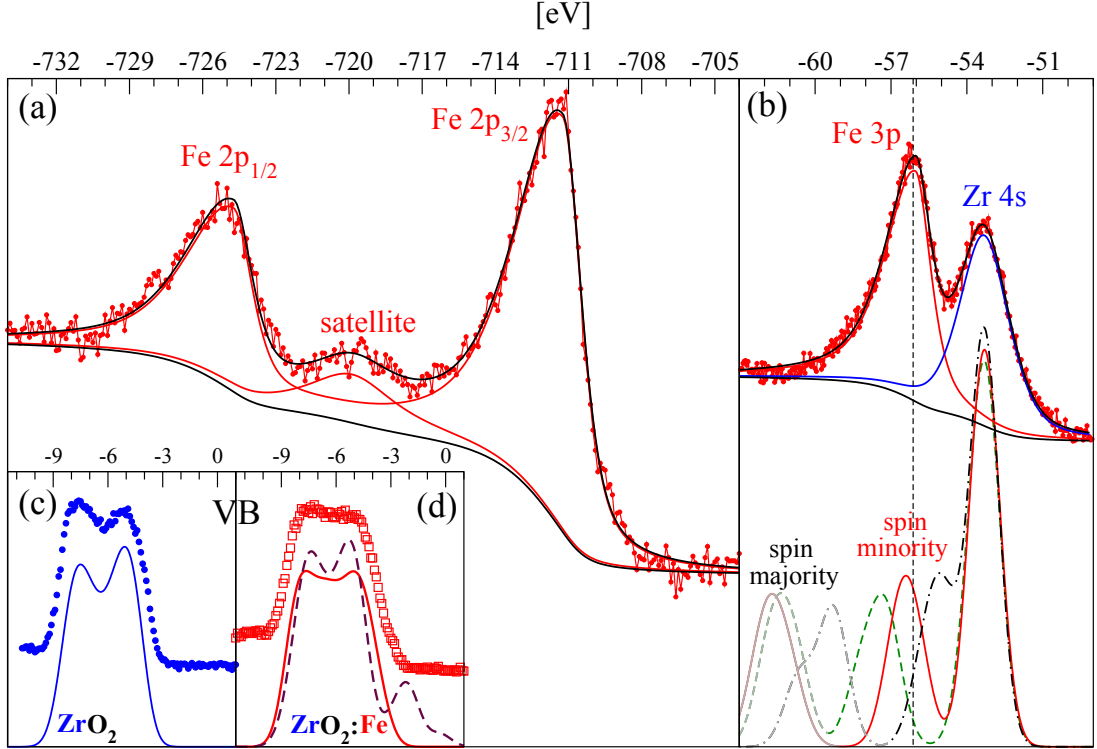}
\caption{(color online)
(a): The $Fe(2p)$ core level photo--emission spectra in $ZrO_2$:$Fe$.
(b): $Fe(3p)$ and $Zr(4s)$ photo--emission spectra and computed 
DOS for $ZrO_2$:$Fe$ 
with $y_{V_O^{^{\bullet\bullet}}/Fe}$ equal to $0$ (green dashed), $0.5$ (red continuous),
$1$ (black dot--dashed). The $Fe(3p)$ majority spin level is in light gray.
(c): Measured and computed valence band (VB) for pure $ZrO_2$.
(d): Measured VB for $ZrO_2$:$Fe$. Computed VB for $ZrO_2$:$Fe$ with $Fe$ doping
substitutional at $y_{V_O^{^{\bullet\bullet}}/Fe}=0.5$ (continuous red line) or interstitials (dashed maroon line).
In panels (b)-(c)-(d) the experimental data (and fit) are vertically shifted respect to the DFT--DOS.
All DOS are obtained at within the GGA. The experimental data were collected with the PHI 5600
instrument (see details in sec.~\ref{sec:Framework_Exp})}
\label{Fig:XPS_and_DFT}
\end{figure}

In Fig.~\ref{Fig:XPS_and_DFT} we report the high resolution spectra
of the $Fe(2p)$ core level (a), the $Fe(3p)$ semi--core (b) levels and the VB (c-d).
In Fig.~\ref{Fig:XPS_and_DFT}.(a-b) the data were fitted with a doublet of asymmetric Voigt functions
for the two main peaks plus a  Voigt function for the satellite on top of a Shirley background
and in Fig.~\ref{Fig:XPS_and_DFT}.(b-d) the spectra are compared with DFT(GGA)--DOS computed as described
in sec.~\ref{sec:Framework_TH}.

The change of the XPS--VB from $ZrO_2$ (blue) to $ZrO_2$:$Fe$ (red)
is in agreement with the DFT(GGA)--DOS obtained considering substitutional iron doping.
In particular experimentally the double peak structure of pure $ZrO_2$
is suppressed with doping. Theoretically this behavior is reproduced only
assuming substitional doping.

\begin{table}[t]
\begin{center}
\linespread{1.}
\parbox{0.8\textwidth}{\caption{Energy distances [$eV$] from $Fe(2p_{3/2})$. \\
                                 Data for iron oxides from Ref.~\onlinecite{Yamashita2008}}
\label{Table:Fe_2p}} \\
\begin{tabular*}{0.8\textwidth}{@{\extracolsep{\fill}} l  c  c  c  c}
\hline
                   & $Fe_2O_3$ & $Fe_3O_4$ & $FeO$       & $ZrO_2$:$Fe$       \\
\hline
$Fe(2p_{1/2})$     &  -13.6    & -13.5     &  -13.6      &  -13.5              \\ 
\hline
satellite          &   -7.8    & not pres. &   -6.0      &   -8.6              \\
\hline
$Fe(3p)$           &  655.4    & not av.   &  653.9      &  655.2
\end{tabular*}
\vspace{-0.5cm}
\end{center}
\end{table}

The core or semi--core levels of TM usually show a
structured shape due to, at least,
four factors: the spin--orbit (SO) splitting, the exchange splitting,
the multiplet splitting and the $eh$ screening to the core--hole.
The SO term is responsible for the
$2p_{1/2}$ - $2p_{3/2}$ splitting $\Delta E_{SO}=13.5$ eV and is not sensitive
to the chemical environment (see Fig.~\ref{Fig:XPS_and_DFT}.(a)).
The exchange and multiplet splitting instead give the characteristic asymmetric shape
of the XPS peaks in metals. Finally the screening
effect, which is strongly sensitive to
the chemical environment~\cite{Huang1995,Takahashi2010,See1995}, can create satellites.
For the $Fe(2p)$ core level
the distance between the satellite and the
$Fe(2p_{3/2})$ peak is a marker of the iron oxidation state~\cite{Yamashita2008}.
Also the position of the $Fe(3p)$ peak (Fig.~\ref{Fig:XPS_and_DFT}.(b)) is sensitive to
the $Fe$ chemical environment~\cite{Yamashita2008}.
The comparison with the values of Ref.~\onlinecite{Yamashita2008}, reported in Table~\ref{Table:Fe_2p},
shows that iron is in the $Fe^{+3}$ oxidation state.

According to our DFT results the $Fe$ oxidation state is strongly related to the presence
of $V_O^{^{\bullet\bullet}}$ in the system (see Fig.~\ref{Fig:DFT_DOS}).
To better describe this point we study
the $Fe(3p)$ semi--core levels with first principles simulations.
Indeed the $Fe(3p)$ wave--functions are spatially localized close
to the $Fe(3d)$, which are in valence, and so are very sensitive to the chemical environment.
The energy of the $Zr(4s)$ level is used as a reference to properly align
the experimental XPS levels with the theoretical DOS.

In our approach the SO coupling term was included, both in the pseudo--potentials and in the hamiltonian,
while the multiplet and the exchange splitting were accounted for by the
exchange--correlation (xc) potential. For the $Fe(3p)$ level we found $\Delta E_{SO}\leq 1\ eV$,
while $\Delta E_{xc}\approx 5\ eV$ between the spin
minority and the spin majority which is clearly visible in Fig.~\ref{Fig:XPS_and_DFT}.(b).
This is overestimated by DFT.
In the case of semi--core levels Takahashi et al.~\cite{Takahashi2010} showed that the screening
effects, which are not included in the present approach,
gives a broadening and a shift of the majority spin channel with, possibly, the creation of satellites.
Indeed we can suppose that these effects would correct the overestimated $\Delta E_{xc}\approx 5\ eV$,
shifting the majority--spin energy level close to the minority one
giving a single asymmetric peak with higher intensity as in the experimental case.
However such an approach is beyond the scope of the 
present work. The minority--spin channel instead is less
affected by screening effects retaining the independent--particle
structure with the onset of the spectrum due to absorption from this channel~\cite{Takahashi2010}.
Thus we compared the energy position of the minority DOS with the measured $Fe(3p)$ XPS spectrum.
In our simulations the distance of the $Fe(3p)$ minority peak from the $Zr(4s)$ level, $\Delta E_y$,
is strongly dependent on $y_{V_O^{^{\bullet\bullet}}/Fe}$ with $\Delta E_{y=0}=1.8$, $\Delta E_{y=0.5}=3.1$
and $\Delta E_{y=1}=4.0\ eV$. The value $\Delta E_{y=0.5}$,
i.e. the configuration with iron in the $Fe^{+3}$ oxidation state,
best agrees with the experimentally measured splitting $\Delta E=2.9\ eV$,
in agreement with the conclusion drawn from
Table~\ref{Table:Fe_2p} and in general from sec.~\ref{sec:results_theory}.


\begin{figure}[t]
\includegraphics[height=0.4\textheight]{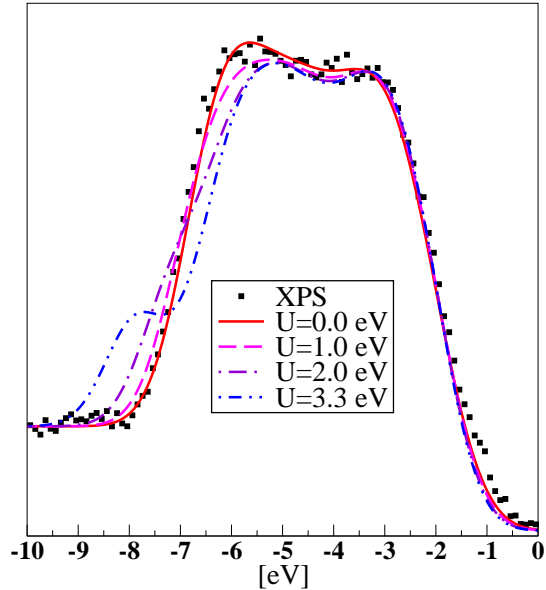}
\caption{(color online)
Valence band of $ZrO_2$:$Fe$. The GGA+$U$ scheme at $x_{Fe}=18.75\ \%$, 
$y_{V_O^{^{\bullet\bullet}}/Fe}=0.5$, for the values of $U=0.0,\ 1.0,\ 2.0,\ 3.3$,
is compared against experimental data. The smearing parameter
used for the plot is $0.06\ Ry$. The experimental data were collected with the PHI 5600
instrument (see details in sec.~\ref{sec:Framework_Exp}). }
\label{Fig:GGA_vs_GGA+U}
\end{figure}

In sec.~\ref{sec:results_theory} we showed
that the electronic properties and in particular the shape of the valence band 
could be strongly influenced by the on--site electronic correlation, by 
the comparison of GGA and GGA+$U$ predictions at $U=3.3\ eV$. 

To decide which of the two scenarios, GGA or GGA+$U$, better describes the experimental situation
we compared the obtained DOS for the charge--compensated case with the measured photo--emission from the VB;
the values $U=0.0,\ 1.0,\ 2.0,\ 3.3\ eV$ are here considered.
To mimic the experimental spectrum, we have superimposed to the 
DFT--DOS a Shirley like background, i.e. a background proportional to the integral of the 
DOS. Also, for a quantitative comparison, we computed the theoretical DOS for
$y_{V_O^{^{\bullet\bullet}}/Fe}=0.5$ and $x_{Fe}=18.75 \%$,
which is the theoretical value closest to the experimental measured doping.

In Fig.~\ref{Fig:GGA_vs_GGA+U} we see that the structure which identifies
the $d$ levels in the GGA+$U$, at the reference value $U=3.3\ eV$, is not present experimentally
and the agreement between theory and experiment is much better in the standard GGA (i.e. $U=0.\ eV$).
At the intermediate values $U=1.0\ eV$ and $U=2.0\ eV$ such structure
is not visible, however the agreement with the experimental results is worse than for the
$U=0.\ eV$ case.
We can conclude that the value $U=0$ best agrees with the photo--emission VB, and that,
given the experimental resolution, the optimal choice of $U$ must be between 0 and 1 $eV$.
Thus in $ZrO_2$:$Fe$ the effect of the self--interaction of the $d$ orbitals, which is
corrected by the Hubbard $U$ term, is smaller than in common iron oxides. This is an
``a posteriori'' justification of the results obtained, in the present work, within 
the GGA. 

\section{Conclusions}

In conclusion we studied iron doped zirconia both theoretically, with first--principles simulations,
and experimentally, with structural, chemical and electronic characterization of thin films
grown by atomic layer deposition.

As expected from simple considerations, iron was found experimentally in the $Fe^{+3}$
oxidation state. We also found that it induces a monoclinic to tetragonal phase transition.
Theoretically the oxidation state was related to presence of
oxygen vacancies which play a key role in the structural phase transition. 
The theoretical findings have been tested with a detailed comparison against
photo--emission spectra of the samples grown by atomic layer deposition to validate the assumptions.
These results are a confirmation that iron doped zirconia could be a good candidate
in view of oxygen sensing applications as reported in the past.

Moreover the presence of vacancies is seen not only to influence the structure of the system but,
theoretically, also to determine the density of states at the Fermi level and the
eventual presence of impurity states in the gap which could be associated to magnetic polarons.
In particular, we discussed how the ratio between oxygen vacancies and the iron atoms
concentration shifts the Fermi level of the system. We found that in the most stable
configuration, the $Fe^{+3}$ iron atoms are charge--compensated by the presence of oxygen
vacancies with a ration of 0.5, i.e. one vacancy each two iron atoms. The resulting
system is a semi--conductor with no impurity state in the gap.

These results should be considered for a correct description of the behavior of
iron doped zirconia, or more in general of high--k oxides doped with valence $+3$
elements, in resistive switching devices. Moreover the absence of impurity states rules
out the magnetic polaron model as a possible mechanism to explain the magnetic properties 
of the system.

Finally we have explored the importance of the Hubbard U correction.
Indeed, theoretically, varying the value of
U from $0\ eV$ to $3.3\ eV$ the electronic propertes of the system change significantly.
We showed that in iron doped zirconia the value $U\approx 0\ eV$ best agrees
with the experimental data, thus indicating that the on site electronic correlation
is low in this system.

\section*{Aknowledgments}
This work was funded by the Cariplo Fundation through the OSEA project 2009-2552.
D.S. and A.D. would like to acknowledge G. Onida and the ETSF Milan node
for the opportunity of running simulations on the ``etsfmi cluster'', P. Salvestrini for technical
support on the cluster and A. Molle and S. Spiga for useful comments and discussions.
We also acknowledge computational resources provided under the project MOSE by CASPUR.


\end{document}